\documentclass[aps,prl,twocolumn,floatfix,showpacs]{revtex4}
\usepackage{graphicx}

\begin{document}

\title{Dynamics of fluctuations below a stationary bifurcation to electroconvection in the planar nematic liquid crystal N4} 

\author{Xin-Liang Qiu and Guenter Ahlers}

\affiliation{Department of Physics and iQUEST,\\ University of
California, Santa Barbara, CA 93106}
\date{\today}
\begin{abstract}
We fitted $C({\bf k},\tau,\epsilon) \propto exp[-\sigma({\bf k},\epsilon)\tau]$ to time-correlation functions $C({\bf k},\tau,\epsilon)$  of  structure factors $S({\bf k}, t, \epsilon)$ of shadowgraph images of fluctuations below a supercritical bifurcation at $V_0 = V_c$ to electro-convection of a planar nematic liquid crystal in the presence of a voltage $V = \sqrt{2}V_0 cos(2\pi f t) $[${\bf k}= (p,q)$ is the wave vector and $\epsilon \equiv V_0^2/V_{c}^2 - 1$]. There were stationary oblique (normal) rolls at small (large) $f$. Fits of a modified Swift-Hohenberg form to $\sigma({\bf k},\epsilon)$ gave $f$-dependent critical behavior for the minimum decay rates $\sigma_0(\epsilon)$ and the correlations lengths $\xi_{p,q}(\epsilon)$.
\end{abstract}


\maketitle

Critical opalescence is a familiar phenomenon that occurs, for instance, near liquid-gas critical points and near the phase separation of binary mixtures.\cite{St71} It is caused by refractive-index fluctuations due to droplets of fluid of lesser or greater than average density or concentration. The fluctuations are the response of the system to the thermal noise inherent in the Brownian motion of the molecules or atoms. The droplet size increases in proportion to a correlation length $\xi$ that diverges at the critical point. Likewise, the decay rate $\sigma$ of the droplets vanishes  as the critical point is approached. When $\xi$ is sufficiently large, ambient light is scattered and the fluid becomes ``milky". Analogous phenomena occur in many spatially extended nonlinear systems as they are driven away from equilibrium and toward an instability, or bifurcation.\cite{RRTSHAB91,WAC95,SAHR00,SA02,AO03,OA03,OOSA04} For those cases the length and time scales are more nearly macroscopic and the fluctuations can be observed directly and in detail with modern shadowgraph techniques.\cite{RHWR89,BBMTHCA96,TC02} Here we present experimental results for $\xi$ and $\sigma$ for electro-convection (EC) of a nematic liquid crystal (NLC). As we will show below, our data are not easily understood by analogy to known equilibrium critical phenomena \cite{St71,Ma76}. 

A NLC consists of elongated molecules that  tend to align locally relative to each other. \cite{Bl83} The alignment direction is called the director $\hat n$. When a NLC is confined between parallel glass plates  with a small spacing $d$ between them, $\hat n$ is influenced by the interaction of the molecules with the glass surfaces. We used surfaces that caused more or less uniform alignment in a unique direction parallel to the plates (planar alignment). This breaks the rotational invariance usually found in isotropic liquids, for instance in Rayleigh-B\'enard convection, \cite{SH77,HS92,OA03} or in NLC with homeotropic alignment ($\hat n$ perpendicular to the plates) \cite{ZA04}.We applied a voltage $V = \sqrt{2}V_0\cos(2 \pi f t)$ between transparent indium-tin oxide electrodes on the inner surfaces of the confining plates. For a NLC with suitable properties this induces a transition (or bifurcation) at $V_0 = V_c$ from a spatially uniform state without flow to a convecting state of lower symmetry. \cite{BZK88} The bifurcation is supercritical, analogous to a critical point in a two-dimensional equilibrium system, e.g. to a Curie point of a ferromagnet. Here we report on the effect of fluctuations on the nature of the bifurcation to EC. 

When the fluctuations are small, one may neglect interactions between them and discuss the phenomena in terms of a linear theory (LT).\cite{LT,GM,Ma76} In that case we expect that $\xi \sim |\epsilon|^{-\nu}$ with $\nu = 1/2$ and $\sigma \sim |\epsilon|^{\lambda}$ with $\lambda = 1$. Here $\epsilon \equiv V_0^2/V_c^2-1$. When the fluctuations become larger, nonlinear terms in the governing equations lead to interactions between them. In that case several universality classes of critical behavior exist for equilibrium systems. They are determined primarily by the dimensionality of the system and by the number of components $n$ of the order parameter. For two-dimensional equilibrium systems one has primarily Ising systems ($n = 1$) and $X-Y$ systems ($n = 2$). It is not certain that this classification carries over in direct analogy to non-equilibrium cases.

The range of $\epsilon$ over which fluctuation interactions become important depends on the coupling of the thermal noise to the system. For EC, the relevant parameter is the ratio of the thermal energy to a characteristic elastic energy of deformation $F = k_BT/\bar k d$ ($\bar k$ is an average elastic constant of the NLC).\cite{RRTSHAB91} On the basis of a Ginzburg-like criterion\cite{Gi60} one can estimate that the critical region is encountered near $\epsilon_c \simeq F^{2/3}$. For the NLC used by us one finds $\bar k \simeq 10^{-11}$ N \cite{FSD03} and thus, for a cell with $d \simeq 25~\mu$m, one has $\epsilon_c \simeq 6\times 10^{-4}$. For much larger $\epsilon$ the prediction of linear theory should pertain.

\begin{figure}
\includegraphics[height=3.0in, width=3.0in, keepaspectratio=true]{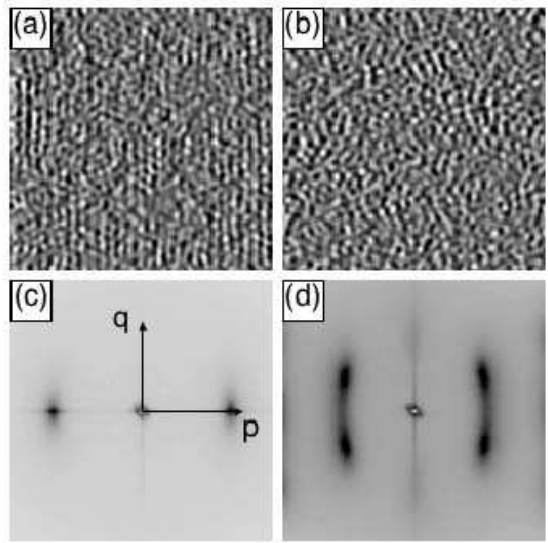}
\vskip 0.03in
\caption{(a,b): Shadowgraph images of size $0.88\times0.88$ mm$^2$ for  $\epsilon = -1.1 \times 10^{-3}$. (a): $f = 4000 Hz$. (b): $f = 25 Hz$.  (c, d): the central part $-7.2 \leq p,q \leq 7.2$ of the averaged structure factor $S({\bf k})$ based on 4096 images for $\epsilon = -1.1 \times 10^{-3}$. (c): $f = 4000 Hz$. (d): $f = 25 Hz$. The director is horizontal.}
\label{fig:Fig1} 
\end{figure}

We made shadowgraph measurements at various values of $f$. We note that a priori there is nothing to indicate that $\epsilon_c$ or the universality class should depend on $f$. We used the same $0.88\times 0.88$ mm$^2$ area of the same sample for all measurements, thus reducing the possibility that a frequency dependent result could actually be due to different sample preparation. As shown in Fig.~\ref{fig:Fig1}a and c,  at large $f = 4$ kHz we found so-called ``normal" stationary convection rolls in the fluctuations below and the patterns above onset. Their wave vectors are parallel to $\hat n$.  For the fluctuations we found $\lambda \simeq 1$, consistent with LT. However, the correlation lengths gave $\nu = 0.40\pm 0.03$, which differs from $\nu = 1/2$ of LT. One might argue that this measured exponent really is an effective exponent representing data in a crossover region from linear to critical behavior. However, then one would have to argue either that the crossover region is different for $\xi$ and $\sigma$ or that the value of $\lambda$ is not altered measurably by fluctuation interactions.

At small $f = 25$ Hz we find that the fluctuations, as well as the patterns above onset, correspond to oblique stationary rolls with their wave vectors forming an oblique angle $|\theta| > 0$ with $\hat n$ (Fig.~\ref{fig:Fig1}b and d). For this case there are two degenerate modes, known as ``zig" and ``zag", with angles $\pm \theta$. Here the critical behavior was qualitatively different. No power law could describe the results for $\sigma_0$ and $\xi$. Both quantities, although they initially changed upon approaching $V_c$, deviated from a power law for $|\epsilon| \alt 0.02$ and approached a  constant value rather than diverging. We find it difficult to reconcile these results with any known critical behavior of equilibrium systems.
   
We took shadowgraph images \cite{RHWR89,BBMTHCA96,TC02} $\tilde{I}_i({\bf x},\epsilon)$ [${\bf x} = (x,y)$ are the spatial coordinates] of a commercial cell \cite{EHC} with a nominal spacing of $d = 25~\mu$m and filled with the NLC Merck N4 (a eutectic mixture of two azoxy compounds, CH3O-C6H4-NO=N-C6H4-C4H9 and CH3O-C6H4-N=NO-C6H4-C4H9) doped with $0.1 \%$ by weight of tetra butylammonium bromide. The alignment was planar. The conductance  was measured at $30^o$C, at $f=40$ Hz and $V_0 = 2.0$ Volt. It was near $6.1 \times 10^{-7} (\Omega m)^{-1}$. We used the total shadowgraph power $P$ to determine that the bifurcation to EC was non-hysteretic and stationary over the range $25 \leq f \leq 8000$ Hz. The power above $V_c$ was used to find $V_c = 6.04~ (11.49)$ Volt for $f = 25 ~(4000)$ Hz. Above the bifurcation the pattern consisted of stationary zig and zag (normal) rolls for $f < f_L \simeq 1.3$ kHz ($f > f_L$). At each voltage we waited 300 s, and then took 4096 shadowgraph images 0.3 s to 1 s apart, depending on $\epsilon$. 

From each image we calculated $I_i({\bf x},\epsilon) \equiv \tilde{I}_i({\bf x},\epsilon) / \tilde{I}_0({\bf x},\epsilon) - 1$. Here  $\tilde{I}_0({\bf x},\epsilon)$ is a background image obtained by averaging 4096 images at the same $\epsilon$. For each $I_i({\bf x},\epsilon)$ we derived the structure factor (the square of the modulus of the Fourier transform) $S_i({\bf k},\epsilon)$ and averaged 4096 $S_i({\bf k},\epsilon)$ to get averaged structure factors $S({\bf k},\epsilon)$, where ${\bf k} = (p,q)$ is the wave vector. 
Figure \ref{fig:Fig1}(c) and (d) show examples. At low frequencies [(d), $f = 25$ Hz] two pairs of peaks correspond to two sets of rolls oriented obliquely to the director. At high frequency [(c), $f = 4$ kHz] only one pair of peaks corresponding to normal rolls was observed. 

\begin{figure}
\includegraphics[width=2.5in]{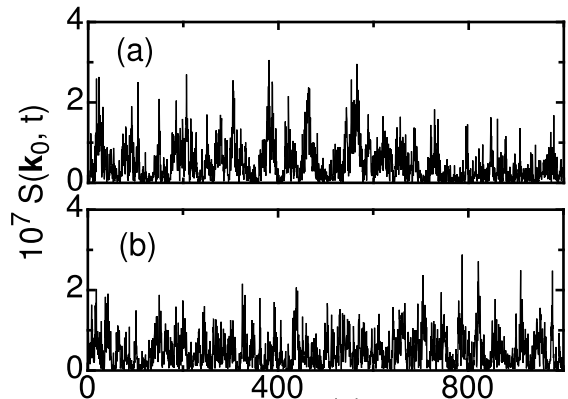}
\vskip 0.03in
\caption{Time series of $S({\bf k_{0}},\epsilon, t)$ at $\epsilon = -1.0 \times 10^{-2}$ for  (a) $f = 25$ Hz, ${\bf k_{0}} = (3.936, 1.968)$ and (b) $f = 4000$ Hz, ${\bf k}_{0} = (4.830, 0)$.}
\label{fig:Fig2}
\end{figure}

Figure \ref{fig:Fig2} shows time series of $S({\bf k_{0}},\epsilon, t)$ for $\epsilon = -0.010$  at ${\bf k} = {\bf k_{0}}$ where $S({\bf k},\epsilon)$ has a maximum. The upper (lower) figure shows the result at low (high) frequency. The fluctuations of the structure factor are random and of similar size at both frequencies.

\begin{figure}
\includegraphics[width=2.5in]{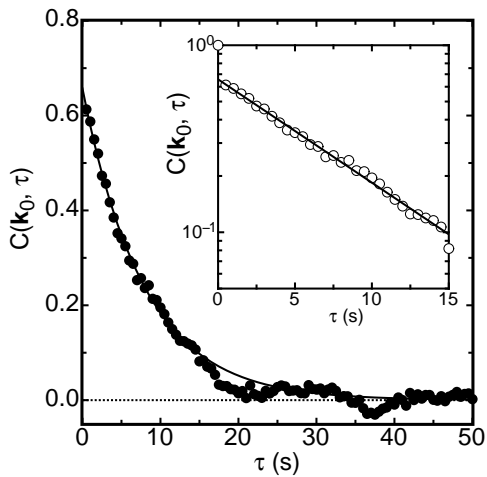}
\vskip 0.03in
\caption{The time auto-correlation function of the fluctuations of the structure factor at ${\bf k_{0}} = (3.936, 1.968)$ for $f = 25$ Hz and $\epsilon = -1.0\times 10^{-2}$. }
\label{fig:Fig3}
\end{figure}

In Fig. \ref{fig:Fig3} we show the auto-correlation function $C({\bf k},\tau) = <S({\bf k},t)S({\bf k},t+\tau)>/<S({\bf k},t)S({\bf k},t)>$ ($<...>$ indicates an average over $t$) of the fluctuations shown in Fig. \ref {fig:Fig2}(a) on linear scales. The insert gives the same data on a semi-logarithmic plot. One sees that the correlations decay exponentially in time. A fit of $Ae^{-\sigma \tau}$ to the data yields $\sigma=0.14$ s$^{-1}$. Although $S({\bf k})$ differs from the structure factor of the refractive index by a $k$-dependent factor [the shadowgraph transfer function ${\cal T}(k)$],\cite{TC02}  we note that $C({\bf k},\tau)$, and thus $\sigma$, are not influenced by ${\cal T}(k)$.

\begin{figure}
\includegraphics[width=3in]{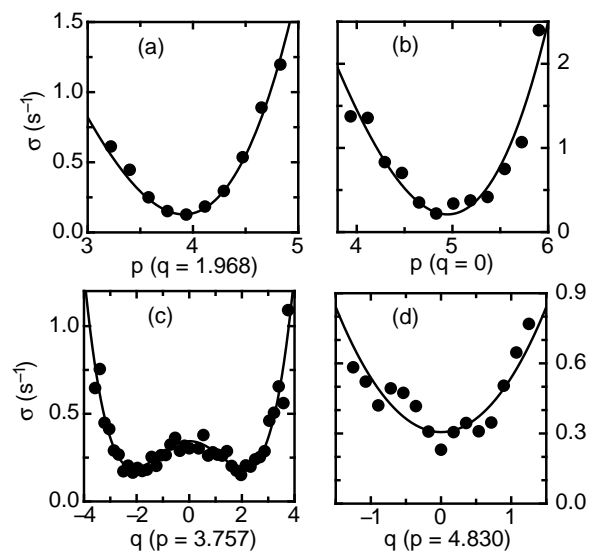}
\vskip 0.03in
\caption{The decay rates $\sigma(p, q, \epsilon)$ as a function of $p$ and $q$ at $\epsilon = -1.0\times 10^{-2}$ for (a,c) $f = 25$ Hz and (b,d) $f=4000$ Hz. The solid lines are fits to the data that yielded  (a) $\sigma_{0} = 0.13, \xi_{p}= 0.37$, (b) $\sigma_{0} = 0.21, \xi_{p}= 0.29$, (c) $\sigma_{0} = 0.18, \xi_{q}= 0.21$, and (d) $\sigma_{0} = 0.31, \xi_{q2}= 0.87, \xi_{q4} = 0.11$.}
\label{fig:Fig4}
\end{figure}

Decay rates of fluctuations at $\epsilon = -0.010$ are shown in Fig.~\ref{fig:Fig4} for oblique (a, c, $f = 25$ Hz)  and normal (b, d, $f = 4$ kHz) rolls along two lines parallel to the $p$ and $q$ axes and passing through the maxima of  $S({\bf k})$ at ${\bf k}_{0}$. In the $p$ ($q$) direction and for $f = 25$ Hz $\sigma$ is shown in Fig.~\ref{fig:Fig4}a (c). 
The Swift-Hohenberg forms
\begin{equation}
\sigma(p,q_0) = \sigma_{0}(\epsilon)\times[(\xi_{p}^{2}(p^{2}-p_{0}^{2})^{2}+1]
\label{eq:p}
\end{equation}
with $q_0 = 1.968$ and 
\begin{equation}
\sigma(p_0,q)=\sigma_{0}(\epsilon)\times[\xi_{q}^{2}(q^{2}-q_{0}^{2})^{2}+1]
\label{eq:q}
\end{equation}
with $p_0 = 3.757$ were fitted to the data (solid lines). 
From Fig.~\ref{fig:Fig4}c it is clear that a simple quadratic form, restricted to either positive  or negative $q$, would not provide a good fit. 

The large-$f$ results for $\sigma$ are shown in Figs.~\ref{fig:Fig4}b and d for the $p$ and $q$ direction respectively. Again, we fitted Eq.~\ref{eq:p} to the data in Fig.~\ref{fig:Fig4}b (solid line).  However, for normal rolls we fitted the function \cite{TK94}
\begin{equation}
\sigma(p = 0,q)= \sigma_{0}(\epsilon)\times(\xi_{q2}^{2}q^{2}+\xi_{q4}^{2}q^{4}+1)
\end{equation}
to the data (solid line).

\begin{figure}
\includegraphics[width=2.5in]{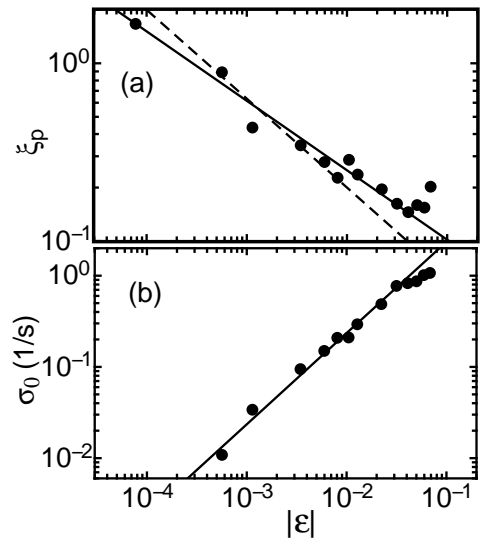}
\vskip 0.03in
\caption{(a) The correlation length $\xi_p$ and (b) the decay rate $\sigma_0$ versus $|\epsilon|$ for $f = 4$ kHz derived from fits of Eq.~\ref{eq:p} to $\sigma(p,q=0)$. The dashed line in (a) corresponds to $\nu = 1/2$. The solid lines are fits to the data that gave (a) $\nu = 0.40\pm 0.03$ and (b) $\lambda = 0.92\pm 0.06$.}
\label{fig:Fig5}
\end{figure}

Each fit yielded a value at a given $\epsilon$ of the minimum decay rate $\sigma_{0}(\epsilon)$, and of the correlation length $\xi_{p}(\epsilon)$ or $\xi_{q}(\epsilon)$. 
For  $f = 4$ kHz we plot $\xi_p$ and $\sigma_{0}$ as a function of $|\epsilon|$  in Fig. \ref{fig:Fig5}a and b respectively. Fits of the results for $|\epsilon| < 0.05$ to power laws yielded $\nu = 0.40 \pm 0.03$ and $\lambda = 0.92 \pm 0.06$. The result for $\lambda$ is reasonably consistent with the prediction of linear theory and suggests either that the critical region has not been reached in this experiment, or that fluctuation interactions do not alter $\lambda$ by a measurable amount. The result for $\nu$ differs significantly from the LT prediction. Of course one could argue in this case that the values of $|\epsilon|$ are too {\it large} for LT to apply; but in that case the range of applicability would have to be different for $\xi_p$ and $\sigma$. Thus the data suggest, but do not firmly establish, a deviation of the critical behavior from the LT prediction.

\begin{figure}
\includegraphics[width=2.5in]{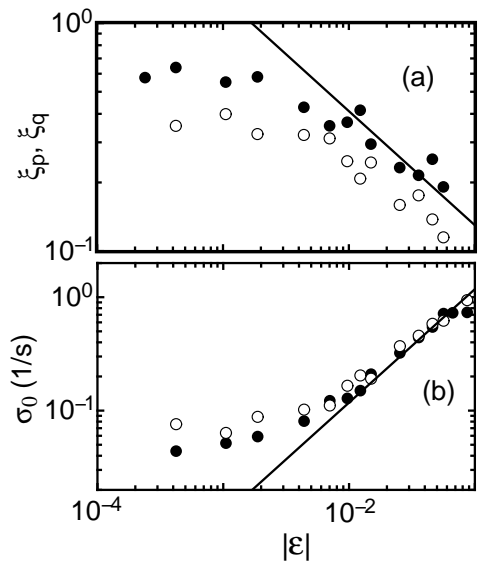}
\vskip 0.03in
\caption{(a) The correlation lengths $\xi_p$ (solid circles) and $\xi_q$ (open circles), and (b) the decay rates $\sigma_0$ derived from fits of Eq.~\ref{eq:p} to $\sigma(p,q_0)$ (open circles) and Eq.~\ref{eq:q} to $\sigma(p_0,q)$ (solid circles), versus $|\epsilon|$ for $f = 25~Hz$. The solid lines correspond to $\nu = 1/2$ and $\lambda = 1$ in (a) and (b) respectively..}
\label{fig:Fig6}
\end{figure}

In Figs.~\ref{fig:Fig6}a and b we show analogous results for $f = 25$ Hz. Here the situation is quite different. For $|\epsilon| \alt 0.02$ deviations from the LT prediction are noticeable for $\sigma_0, \xi_p$, and $\xi_q$. Indeed, a power law can not be used to fit the data. The results suggest that all three parameters approach a finite value at $\epsilon = 0$. This behavior differs from that of any known equilibrium system near a critical point.  However, it would be consistent with a first-order phase transition in the presence of strong noise.

We also measured the total power $P = \int_0^\infty S({\bf k}) d^2k$  at various $|\epsilon|$, and found that  the $|\epsilon|$ dependence of $P$ depended on $f$. Correlation lengths derived from the widths of $S({\bf k})$ agreed with those reported here. At small $f$ we also found that the maximum of $S({\bf k})$ remained finite, whereas one would have expected it to diverge at $\epsilon = 0$.  Those results also differed from previous measurements for EC near a Hopf bifurcation in the NLC I52. \cite{SAHR00,SA02} This work will be reported in a more detailed publication.

In this Letter we presented measurements of the dynamics of thermally driven fluctuations near the bifurcation to stationary electro-convection of the nematic liquid crystal N4. The results for the decay rate $\sigma_0$ and the correlation lengths $\xi_p$ and $\xi_q$ for oblique rolls at small driving frequency $f$ tend toward a constant value at the bifurcation. This differs from the prediction of linear theory which yields power-law singularities with exponents $\lambda = 1$ and $\nu = 1/2$ respectively. It also differs from known possibly relevant equilibrium critical  behavior which yields vanishing decay rates and diverging correlation lengths, albeit with modified exponent values. It is consistent with a first-order transition in the presence of strong noise where no hysteresis would be observed. The results for normal rolls at large $f$ are more nearly consistent with LT, although the value of $\nu$ is somewhat lower than the predicted value 1/2. There was no a priori reason to expect a dependence of the results on the driving frequency.
 
One of us (GA) is grateful to P.C. Hohenberg and W. Pesch for numerous stimulating discussions. This work was supported by the US National Science Foundation through Grant  DMR02-43336.


\begin{references}
\bibitem{GM} From the viewpoint of critical phenomena the results are equivalent to those for the Gaussian Model (Ref.\cite{Ma76}).
\end{references}
\end{document}